\documentclass[a4paper]{jpconf}
\usepackage{graphicx}
\pdfoutput=1
\usepackage{listings,multicol}
\usepackage{parcolumns}
\usepackage{xcolor}

\definecolor{dkgreen}{rgb}{0,0.6,0}
\definecolor{gray}{rgb}{0.5,0.5,0.5}
\definecolor{mauve}{rgb}{0.58,0,0.82}
\lstdefinestyle{myScalastyle}{
  frame=tb,
  float=*,
  language=scala,
  aboveskip=3mm,
  belowskip=3mm,
  showstringspaces=false,
  columns=flexible,
  basicstyle={\small\ttfamily},
  numbers=none,
  numberstyle=\tiny\color{gray},
  keywordstyle=\color{blue},
  commentstyle=\color{dkgreen},
  stringstyle=\color{mauve},
  frame=single,
  breaklines=true,
  breakatwhitespace=true,
  tabsize=3,
}

\begin{document}
\title{“Big Data” in HEP: A comprehensive use case study}

\author{Oliver Gutsche$^1$, Matteo Cremonesi$^1$, Peter Elmer$^2$, Bo Jayatilaka$^1$, Jim Kowalkowski$^1$, Jim Pivarski$^2$, Saba Sehrish$^1$, Cristina Mantilla Suárez$^3$, Alexey Svyatkovskiy$^2$, Nhan Tran$^1$}

\address{$^1$Fermi National Accelerator Laboratory, Batavia, IL, USA}
\address{$^2$Princeton University, Princeton, NJ, USA}
\address{$^3$Fermi National Accelerator Laboratory, Batavia, IL, USA; now Johns Hopkins University, Baltimore, MD, USA}

\ead{gutsche@fnal.gov}

\begin{abstract}
Experimental Particle Physics has been at the forefront of analyzing the world’s largest datasets for decades. The HEP community was the first to develop suitable software and computing tools for this task. In recent times, new toolkits and systems collectively called “Big Data” technologies have emerged to support the analysis of Petabyte and Exabyte datasets in industry. While the principles of data analysis in HEP have not changed (filtering and transforming experiment-specific data formats), these new technologies use different approaches and promise a fresh look at analysis of very large datasets and could potentially reduce the time-to-physics with increased interactivity.
%

In this talk, we present an active LHC Run 2 analysis, searching for dark matter with the CMS detector, as a testbed for “Big Data” technologies. We directly compare the traditional NTuple-based analysis with an equivalent analysis using Apache Spark on the Hadoop ecosystem and beyond. In both cases, we start the analysis with the official experiment data formats and produce publication physics plots. We will discuss advantages and disadvantages of each approach and give an outlook on further studies needed.
\end{abstract}

\section{Introduction}

In 2012, Particle Physics entered a new age. With the discovery of the Higgs Boson, the Standard Model was extended with the missing mechanism that gives rise to particle masses. This theory, developed since the 1960's and constrained by numerous experiments before discovery, followed a predictable path. Now that the Higgs Boson has been discovered, the way forward is wide open. Many different theories that could explain the shortcomings of the Standard Model need to be investigated.

Particle physics has always been at the forefront of analyzing the world's largest datasets. Although we constrain ourselves in this paper to High Energy Physics (HEP), where known particles are made to collide at the highest energies possible, the underlying data organization holds for all sub-fields of particle physics. The most basic concept of how data in HEP is organized is an event: all detector signals associated with a single beam crossing and high-energy collision. Events are the atomic unit of HEP data and may be processed separately, which is why the computational problems of particle physics can be easily parallelized. 

Events must be reconstructed to convert detector signals into measurements of particles produced in collisions. This is usually done centrally by each experiment. The reconstructed events are then input to the final analysis done by individual researchers of groups of researchers exploring a multitude of physics questions. This process is very idiosyncratic, as individual researchers or groups are searching for different physics phenomena in the data. This results in a challenging computational problem, where as many as thousands of physicists analyze the same datasets to extract different physics results.

The analysis process uses properties of the event such as energy or momentum of particles produced in the collisions. It is based on comparing the distribution of properties measured in many events with theoretical predictions of the same property, either calculated empirically or simulated using Monte Carlo techniques. Particle physics is a statistical science. Statistically significant numbers of recorded and simulated events are required to make claims.

The particle physics community was the first to develop suitable software and computing tools for this task. Since the vast majority of collisions result in uninteresting events and the rate is enormous, this community also developed procedures to avoid recording all particle collisions, collectively called the trigger. In this first pass, quantities of interest are calculated quickly to decide whether to save or discard the event in real time.

Trigger selections are as coarse as possible, so that analysis selections are likely to be their subsets. Each physics group analyzes their data by filtering unnecessary events (which we call "skimming") and eliminating unnecessary variables from the communal dataset (which we call "slimming") to get a manageable sub-sample. This sub-sample might be further filtered and transformed to optimize the statistical significance of the measurement or possible discovery. Understandably, this is an iterative process, which must be repeated in response to discoveries and mistakes.

With ever increasing data volumes in particle physics, this process gets harder to perform interactively, simply due to the time required to read, filter, and transform the data. In the future, exponentially growing datasets will make this problem acute. New techniques will be required if we are to continue exploring the nature of matter and the universe.

Recently, new toolkits and systems have emerged outside of the HEP community to analyze Petabyte and Exabyte datasets in industry, collectively called "Big Data." These new technologies use different approaches and promise a fresh look at analysis of extremely large datasets. In this paper, we present an active LHC Run 2 analysis, searching for dark matter with the CMS detector, as a testbed for the application of “Big Data” technologies to a HEP analysis problem. We directly compare the traditional analysis with an equivalent analysis performed in Apache Spark~\cite{Zaharia:2010:SCC:1863103.1863113}. In both cases, we start the analysis with the communal dataset and produce publication physics plots using traditional and Big Data technologies. 

\section{The Physics Use Case}

Today, we know that the universe consists primarily of matter that we do not understand. Only 4--5\% of the universe is made of matter described by the Standard Model~\cite{cosmology}. The Standard Model describes 12 elementary particles: 3 leptons (electron, muon, tau), 3 neutrinos (electron-neutrino, muon-neutrino, tau-neutrino), and 6 quarks (up, down, strange, charm, bottom, top). It describes 4 force carriers or bosons: photon, W, Z, and gluons, as well as one Higgs Boson to provide mass to the leptons, quarks, and W/Z bosons. The remainder of the universe consists of (a) something invisible to all of these forces except perhaps the W and Z bosons, which is unlike any known particle and is called Dark Matter (25\%), and (b) something that appears to have negative pressure, called Dark Energy (70\%).

If Dark Matter {\it does} interact with the W or Z bosons (the Weak Force), then it would be produced in particle collisions, along with the visible particles well described by the Standard Model. Dark Matter particle(s) would propagate through the detector without being detected. However, the total momentum of the collision, transverse to the beam, would be preserved, leading to an apparent momentum imbalance among the visible particles, known as missing transverse energy.

The Dark Matter search conducted in this study is performed with the Large Hadron Collider (LHC)~\cite{1748-0221-3-08-S08001} at CERN, Geneva, Switzerland. The LHC is the highest energy particle collider in the world, located in Geneva, Switzerland. It accelerates protons to an energy of 6.5 TeV in two circular, evacuated beam pipes. The beams of protons are brought to collision in four points around the almost 17 mile circumference ring. The Compact Muon Solenoid (CMS) Collaboration built, maintains, and operates the CMS detector~\cite{cms} at one of these collision points. The CMS detector consists of detector components that measure different properties of the particles produced in a collision, such as tracks left by charged particles and energy deposits from all particles that interact via photons and gluons. The detector components are arranged concentrically around the collision, like an onion, to provide the highest detail near the collision point. The dense distribution of sub-detectors compared to other experiments of its type ("compact"), and the detailed instrumentation for measuring muons gives the 14,500 ton detector and the collaboration its name.

In particular, this Dark Matter search is looking for a signature in the events commonly referred to as "mono-X" where "X" can be a light quark or gluon, a vector boson, or a heavy quark such as a bottom or top quark. We focus our search on the “monoTop” signature~\cite{CMS-PAS-EXO-16-017}, where the detectable particle is a single, unbalanced top quark.

\section{The Traditional Analysis Workflow}
\label{sec:root_workflow}

The traditional user analysis workflow for CMS data applies two C++ frameworks to the centrally produced dataset: CMSSW~\cite{cmssw}, specially designed for analyzing CMS data, and ROOT~\cite{root}, which is a general, experiment-independent C++ toolkit. The ROOT framework provides statistical tools and a serialization format to persist reconstructed and transformed objects in files.

Although these C++ frameworks can be very efficient, their organization can be difficult for end-users to understand. Moreover, it operates at a low level of abstraction, and even if a user knows how to process data efficiently, the difficulty in setting up a manual procedure may be outweighed by the time required to do so.

Most data analysts or analysis groups start by translating the class structure of the data into a "flat ntuple," in which events are rows of a table with primitive numbers or arrays of numbers as columns. These ntuples are written to files using the ROOT framework. They may then be analyzed without the CMSSW framework, with minimal dependencies.

This "ntupling" step requires significant resources as it accesses the centrally produced data and MC samples. GRID resources are used to exploit parallelization where workflow management systems take care of orchestrating the whole analysis workflow. Still significant burden of bookkeeping and failure re-submission is put on the individual analyst or analysis groups. 

Often, the ntuples are still too big for interactive analysis (seconds or minutes between operations, so that a data analyst can respond to results without losing his or her train of thought). Most analysis groups therefore introduce additional steps in which the ntuples themselves are skimmed and slimmed.

In the last step of the analysis, quantities from the final ntuple are aggregated and plotted as histograms. In some cases, data from qualitatively different datasets must be combined to make a single plot, for instance by stacking Monte Carlo contributions to the data (to see if the real data exceeds the sum of known processes) or combining Monte Carlo samples with different weights.

\begin{figure}
\begin{center}
\includegraphics[width=0.5\textwidth]{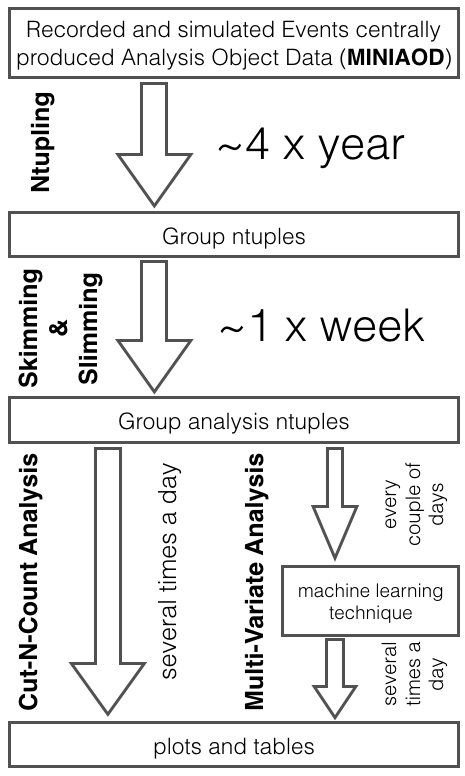}
\caption{\label{fig:workflow}The traditional workflow showing the Ntupling, Skimming and Slimming and final analysis steps.}
\end{center}
\end{figure}

The time scale of the complete Dark Matter workflow, shown in Fig.~\ref{fig:workflow}, can range from days to weeks, depending on how many reconstructed and simulated events are needed for the analysis. The first step is repeated about four times a year and the produced ntuples can be used by more than one analysis. The skimming and slimming step is executed between once and four times a month. The actual analysis step, producing plots, is repeated many times a day, since it represents the iterative and interactive analysis.

From a communal dataset of 200~Terabytes, the analysis ntuples have a size of about 2~Terabytes. These are then skimmed and slimmed down to several Gigabytes that are used for exploration and producing plots.

\section{The Spark Analysis Workflow}
\label{sec:spark_workflow}

The emergence of "Big Data" toolkits and systems in industry opens up new possibilities for analyze petabyte and even exabyte size datasets. In exploring their applicability for HEP analysis, our goals are not only to reduce the time it takes an analyst or analysis group to go from centrally produced input samples for data and MC to publication-quality plots (time-to-physics) in the advent of the exabyte era, but also to educate our graduate students and postdocs to use industry-based technologies and improve their chances on the job market outside of academia and HEP's attractiveness for potential candidates. Tools like the ROOT toolkit are used almost exclusively in the HEP community. Expanding the used techniques to include Big Data technologies would allow to be part of larger communities reaching outside of our field. 

We chose to investigate the Apache Spark toolkit as a starting point. The main goal is to translate the skimming and slimming step of the use case analysis to the Spark framework. We chose this step to simplify the translation problem and to avoid further complications in the ntupling step, which normally requires re-running experiment framework code. This step can be tackled in a later study.

We translated the group ntuples stored in ROOT format into a suitable file format for Spark. For simplicity, we chose the AVRO format~\cite{avro}. This choice might not be optimal but provided a solid starting point for the investigation. We auto-generated the AVRO files from the ROOT files using a newly developed root converter package~\cite{rootconverter}. This package allows the conversion of any complex ROOT file into AVRO and proved to be sufficiently flexible for the used analysis use case. The generated schema for this particular analysis can be found here~\cite{avroschema}.

The actual analysis code was translated from its ROOT based C++ code basis into Scala~\cite{scala}, chosen because of its excellent native support in Spark. Scala allowed to significantly simplify the analysis workflow consisting of two loops over the input. In the case of MC input, first a sum of weights determined in the simulation itself is calculated by looping over all events in the input. This is followed by a second event loop that performs the actual skimming and slimming. The Scala code is shown here:

\begin{lstlisting}[style=myScalastyle]
// Reference the whole dataset (not individual files)
val mcsample = avrordd("hdfs://path/to/mcsample/*.avro")

// Sum of Weights from Simulation: First pass (and cache for later)
mcsample.persist()
val mc_sumOfWeights = mcsample.map(_.GenInfo.weight).sum

// Main Event Selection: Second pass on data in cluster's memory
val result = mcsample.filter(cuts).map(toNtuple(_, mc_sumOfWeights, mc_xsec))

// Save as ntuple
result.toDF().write.parquet("hdfs://path/to/mcsample_ntuple")
\end{lstlisting}

As a development system, We used a 10 node cluster of Intel Xeon CPU E5-2680 v2 @ 2.80GHz CPU processors with 256 GB RAM interconnected through 10 Gigabit Ethernet. The Cloudera~\cite{cloudera} Hadoop distribution (CDH 5.9.1) was used. Spark jobs were scheduled through YARN. Input AVRO files were stored on an HDFS file-system spanning the cluster. Spark was automatically using the whole cluster for the two event loops.

\section{The Usability Comparison}

To compare the traditional ROOT-based workflow with its Spark-based translation, we were looking at the problem from the  "physicist" point of view and we are not assuming any advanced experience with optimizing GRID and High-Throughput (HTC) workflows. We took the current ROOT workflow of the Dark Matter analysis and compared it with the Spark workflow. The ROOT workflow was executed at CERN on lxplus/lxbatch while the Spark workflow used the cluster at Princeton described in Section~\ref{sec:spark_workflow}. In the following, we describe the experience of two beta-users who didn't have experience with either workflow before.

The ROOT workflow consisted of a master script producing a series of secondary scripts to execute the analysis. The first step to calculate the sum of weights was executed serially using an interactive login node. The second step was executed in batch mode on lxbatch, which is using the LSF batch scheduler~\cite{lsf}. This was compared to the Spark workflow consisting of the two steps executed in Scala (see Section~\ref{sec:spark_workflow}). The Spark workflow also persisted the input in memory, although this had to be activated manually. The beta-testers preferred the simplicity of the Spark workflow over the ROOT workflow setup, but also mentioned that the ROOT workflow could have been optimized without a lot of effort. They particularly liked the ease of persisting the input in Spark.

Comparing the analysis code itself, the testers preferred the ROOT analysis code because they were very familiar with C++ and ROOT from their previous analysis experience. Scala was a new language for them and had a learning curve for them compared to C++ and ROOT.

We were particularly interested in comparing the bookkeeping efforts of the two workflows. The testers noted that the ROOT workflow was designed for specific batch systems (in this case LSF). They tried but were not able to move it easily to another batch system. The partitioning of the inputs of the ROOT workflow was handled through a sophisticated suite of shell scripts relying on the physical location of the input files (in this case EOS at CERN~\cite{eos}). This was compared to the Spark workflow, which according to the testers showed very good portability. They were able to move the Spark workflow to CERN/lxplus in a very short time. The partitioning in Spark can use automatic or custom facilities. The testers remarked its ease of use and that it is superior to the ROOT workflow approach. In summary, the bookkeeping effort was significantly reduced in the Spark workflow case and the portability is significantly easier to accomplish.

\section{A First Look at a Performance Comparison}

Performance comparisons are complicated by nature when dealing with sophisticated multi-step workflows using a variety of different computing systems. We are presenting a first attempt of comparing the performance of the ROOT and Spark workflows. We can only extract qualitative conclusions from this first comparison, for a more in-depth comparison more studies have to be performed.

We were using the same single lxplus node at CERN and were restricting the workflow to a single core to use a comparable hardware setup for both workflows. Lxplus is an interactive login cluster, the impact from other users using the same node at the same time could not be excluded but was assumed to be small. 

The workflow setup is described in Sections \ref{sec:root_workflow} and \ref{sec:spark_workflow}. The input was a single ROOT file of 1 GB size, or a corresponding single AVRO file of size 2 GB. Compression was not activated for the AVRO file while the ROOT file was compressed. This is a significant difference as the ROOT file has to be decompressed during reading. This difference is going to be removed in future studies.

Table~\ref{tab:performance} shows the timing measurements performed on the single lxplus node  described above.

\begin{table}[h]
\caption{\label{tab:performance}First timing measurements performed on a single lxplus node using one core. Please note that the input was a compressed ROOT file that was automatically decompressed during reading while the input to the Spark workflow was uncompressed AVRO.}
\begin{center}
\begin{tabular}{lrr}
\br
&\bf Spark&\bf ROOT\\
\mr
Analysis run without caching&9.4 sec&32.7 sec\\
Reading from local disk \& Computation&4.3 sec&26.8 sec\\
Writing to local disk&5.1 sec&5.9 sec\\
\mr
Analysis run with caching&5.5 sec&\\
Reading from memory cache \& Computation&0.4 sec&\\
Writing to local disk&5.1 sec&\\
\br
\end{tabular}
\end{center}
\end{table}

Qualitatively, the execution times show a speed-up of the Spark workflow when reading from memory cache and also suggest that the performance of the Spark workflow is dominated by reading the input and writing the output. The writing of the output is comparable for the ROOT and Spark workflows and the reading of the input of the ROOT workflow shows the impact of decompressing the input. Because the comparison of the two workflows is far from trivial and not straight forward, more work needs to go into making the comparison fair and conclusive. Our conclusion at this point is that Spark is not orders of magnitude slower than ROOT.

\section{Conclusions and Next Steps}

We presented our work on using Apache Spark to reproduce a use-case HEP Dark Matter analysis and to compare to the corresponding ROOT based analysis workflow. We conclude that the initial Spark implementation of the analysis was more user friendly and portable than the ROOT workflow, which didn't come to a great performance cost. We can say qualitatively that the Spark workflow is not orders of magnitude slower than the ROOT workflow and shows promise for the future.

The next steps in investigating Apache Spark for HEP analysis are to develop a fair and conclusive comparison of the Spark and ROOT workflows by looking critically an the architecture of both workflows, and to optimize both workflows within the capabilities of the respective technologies.

We think Spark and other industry-based Big Data technologies have great potential for HEP analysis and we are planning to continue to investigate Apache Spark for HEP analysis and also expand to other industry-based Big Data technologies.

\section{Acknowledgments}

We would like to thank the CMS collaboration and the LHC to provide the data for the use case and the ROOT based workflow. This work was partially supported by Fermilab operated by Fermi Research Alliance, LLC under Contract No. DE-AC02-07CH11359 with the United States Department of Energy, and by the National Science Foundation under Grant~ACI-1450377, Cooperative Agreement PHY-1120138.

\section*{References}
\bibliography{main}

\providecommand{\newblock}{}
\begin{thebibliography}{10}
\expandafter\ifx\csname url\endcsname\relax
  \def\url#1{{\tt #1}}\fi
\expandafter\ifx\csname urlprefix\endcsname\relax\def\urlprefix{URL }\fi
\providecommand{\eprint}[2][]{\url{#2}}

\bibitem{Zaharia:2010:SCC:1863103.1863113}
Zaharia M, Chowdhury M, Franklin M~J, Shenker S and Stoica I 2010 {\em
  Proceedings of the 2Nd USENIX Conference on Hot Topics in Cloud Computing\/}
  HotCloud'10 (Berkeley, CA, USA: USENIX Association) pp 10--10
  \urlprefix\url{http://dl.acm.org/citation.cfm?id=1863103.1863113}

\bibitem{cosmology}
Ade P~e~a 2014 {\em A\&A\/} {\bf 571} A16
  \urlprefix\url{http://dx.doi.org/10.1051/0004-6361/201321591}

\bibitem{1748-0221-3-08-S08001}
Evans L and Bryant P 2008 {\em Journal of Instrumentation\/} {\bf 3} S08001
  \urlprefix\url{http://stacks.iop.org/1748-0221/3/i=08/a=S08001}

\bibitem{cms}
Chatrchyan S~e~a (CMS Collaboration) 2008 {\em JINST\/} {\bf 3} S08004

\bibitem{CMS-PAS-EXO-16-017}
 2016 {Search for dark matter in association with a boosted top quark in the
  all hadronic final state} Tech. Rep. CMS-PAS-EXO-16-017 CERN Geneva
  \urlprefix\url{http://cds.cern.ch/record/2160266}

\bibitem{cmssw}
Elmer P, Hegner B and Sexton-Kennedy L 2010 {\em J. Phys. Conf. Ser.\/} {\bf
  219} 032022

\bibitem{root}
Brun R and Rademakers F 1997 {\em Nuclear Instruments and Methods in Physics
  Research Section A\/} {\bf 389} 81 -- 86 ISSN 0168-9002 new Computing
  Techniques in Physics Research V
  \urlprefix\url{http://www.sciencedirect.com/science/article/pii/S016890029700048X}

\bibitem{avro}
  \urlprefix\url{http://avro.apache.org}

\bibitem{rootconverter}
  \urlprefix\url{https://github.com/diana-hep/rootconverter}

\bibitem{avroschema}
  Avro Schema: mc\_schema.avsc in
  \urlprefix\url{https://github.com/CMSBigDataProject/SparkBaconAnalyzer/blob/master/test/data}

\bibitem{scala}
  \urlprefix\url{http://www.scala-lang.org}

\bibitem{cloudera}
  \urlprefix\url{http://www.cloudera.com/products/apache-hadoop.html}

\bibitem{lsf}
  \urlprefix\url{http://www-03.ibm.com/systems/spectrum-computing/products/lsf/index.html}

\bibitem{eos}
  \urlprefix\url{https://eos.web.cern.ch}

\end{thebibliography}

\end{document}